\documentclass[fleqn,usenatbib]{mnras}

\usepackage{newtxtext,newtxmath}
\usepackage{hyperref}
\usepackage{booktabs}
\usepackage{lscape} 

\usepackage[T1]{fontenc}

\DeclareRobustCommand{\VAN}[3]{#2}
\let\VANthebibliography\thebibliography
\def\thebibliography{\DeclareRobustCommand{\VAN}[3]{##3}\VANthebibliography}

\usepackage{graphicx}	
\usepackage{amsmath}
\usepackage{longtable}
\usepackage{subcaption}
\usepackage{url}
\usepackage{multicol}

\title[GRB Missing Light Curve prediction]{Predicting Missing Light Curves of Gamma-Ray Bursts with Bidirectional-LSTM: An Approach for Enhanced Analysis}

\author[Sourav et al.]{
Shashwat Sourav,$^{1}$\thanks{shashwat20@iiserb.ac.in}
Amit Shukla,$^{1}$\thanks{amit20@iiserb.ac.in}
Rajeev Ranjan Dwivedi,$^{1}$\thanks{rajeev22@iiserb.ac.in}
and
Kartikey Singh$^{1}$
\\
$^{1}$Indian Institute of Science Education and Research, Bhopal, Madhya Pradesh, India\\}

\date{Accepted XXX. Received YYY; in original form ZZZ}

\pubyear{2023}

\begin{document}
\label{firstpage}
\pagerange{\pageref{firstpage}--\pageref{lastpage}}
\maketitle

\begin{abstract}

Gamma-ray bursts (GRB) are powerful transient events that emit a large output of gamma rays within a few seconds. Studying these short bursts is vital for cosmological research since they originate from sources observed at large redshifts. To effectively carry out these studies, it is crucial to establish a correlation between the observable features of GRBs while reducing their uncertainty.
For these reasons, a comprehensive description of the general GRB light curve (LC) would be crucial for the studies. However, unevenly spaced observations and significant gaps in the LC, which are primarily unavoidable for various reasons, make it difficult to characterize GRBs. Therefore, the general classification of GRB LCs remains challenging. In this study, we present a novel approach to reconstruct gamma-ray burst (GRB) light curves using bidirectional Long Short-Term Memory (BiLSTM). Experimental results show that the BiLSTM approach performs better than traditional methods and produces smoother and more convincing reconstructions for GRBs. 

\end{abstract}

\begin{keywords}
cosmological parameters - gamma-ray bursts - methods: data analysis - software: simulations
\end{keywords}



\begin{figure*}
    \centering
    \includegraphics[scale=0.6]{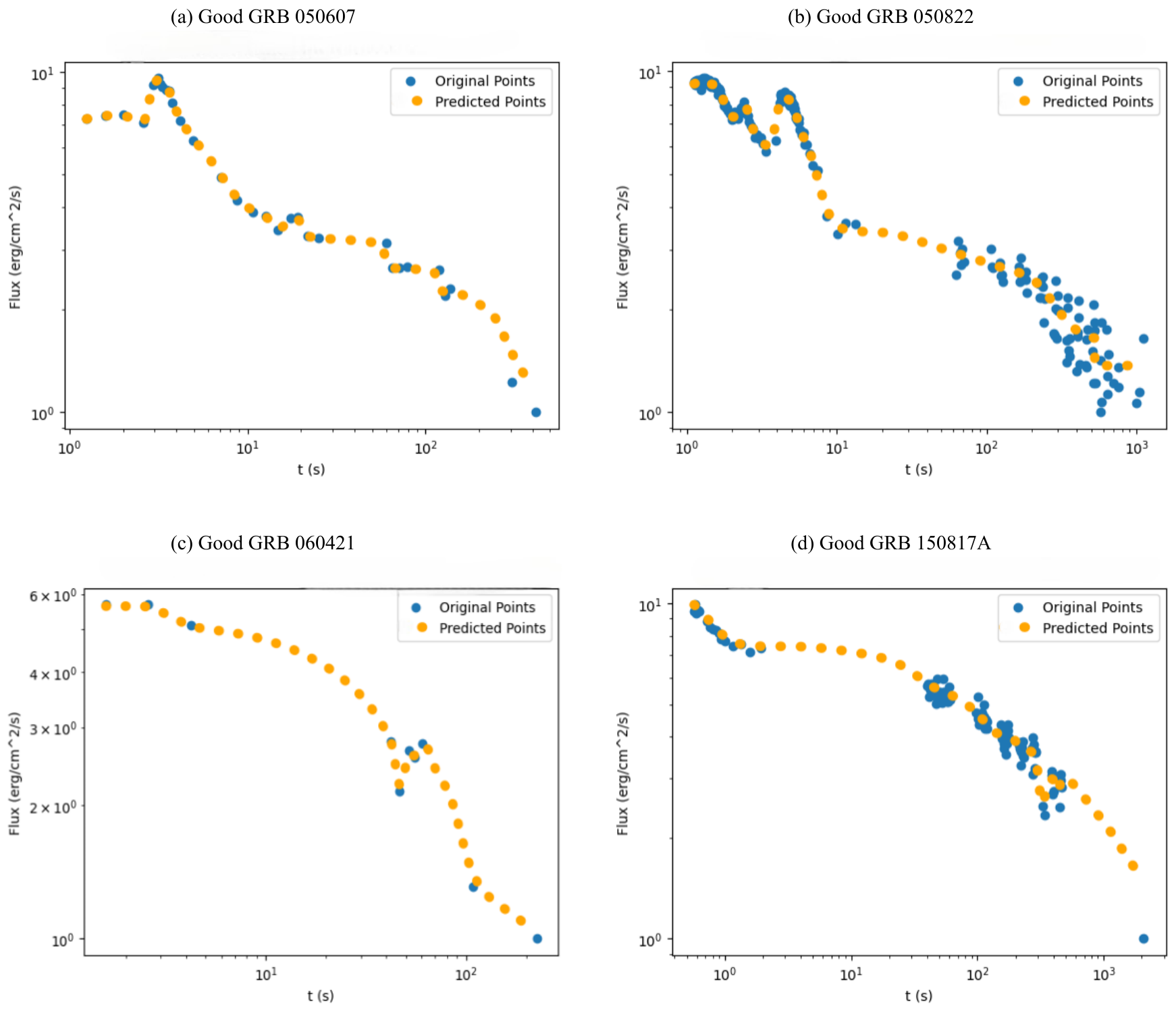}
    \caption{Bi-directional LSTM Light Curve Reconstruction for Good GRBs.}
    \label{fig: Break Bump GRBs1}
\end{figure*}

\begin{figure*}
    \centering
    \includegraphics[scale=0.6]{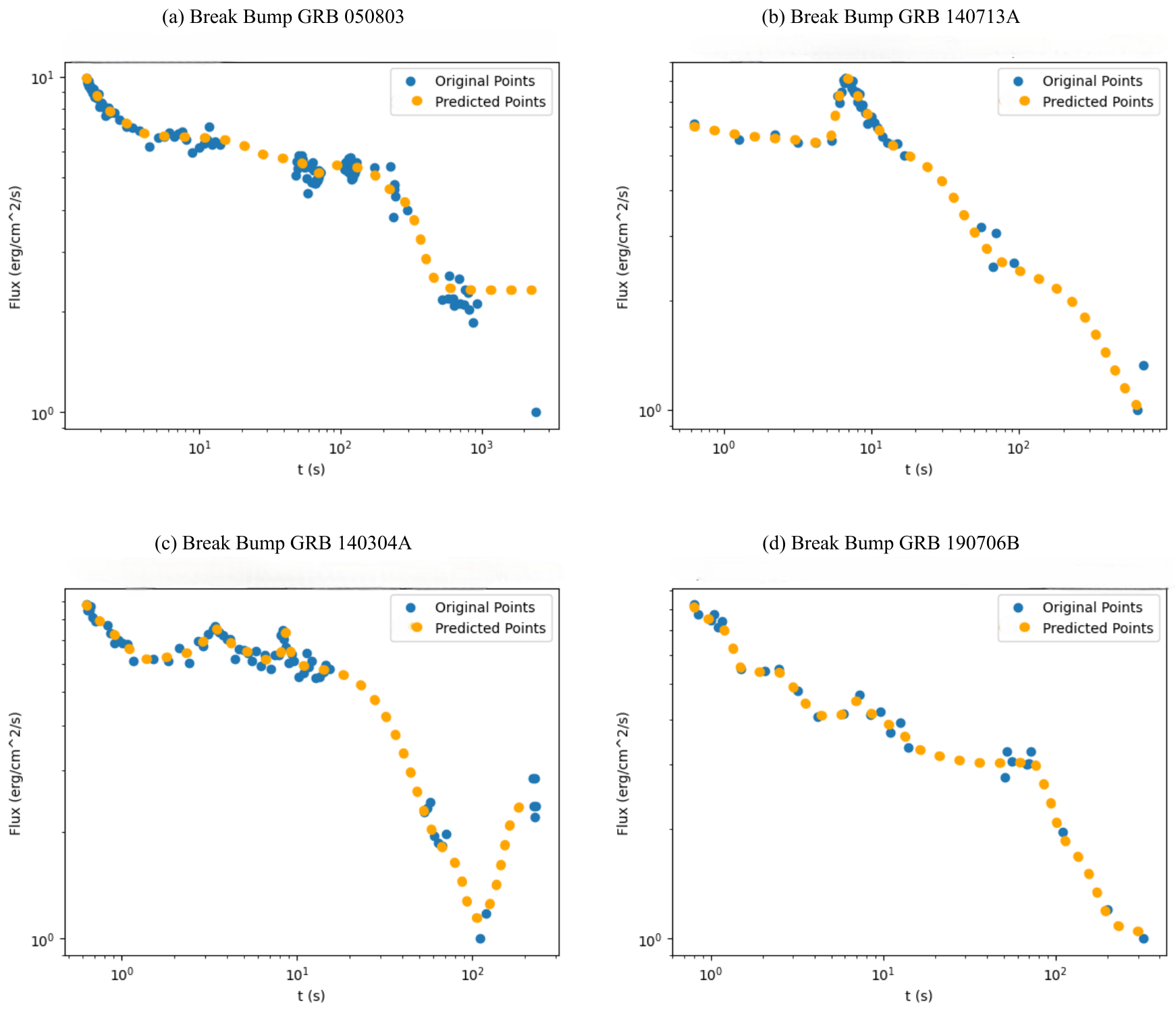}
    \caption{Bi-directional LSTM Light Curve Reconstruction for Break Bump (with a single break) GRBs.}
    \label{fig: Break Bump GRBs2}
\end{figure*}

\begin{figure*}
    \centering
    \includegraphics[scale=0.6]{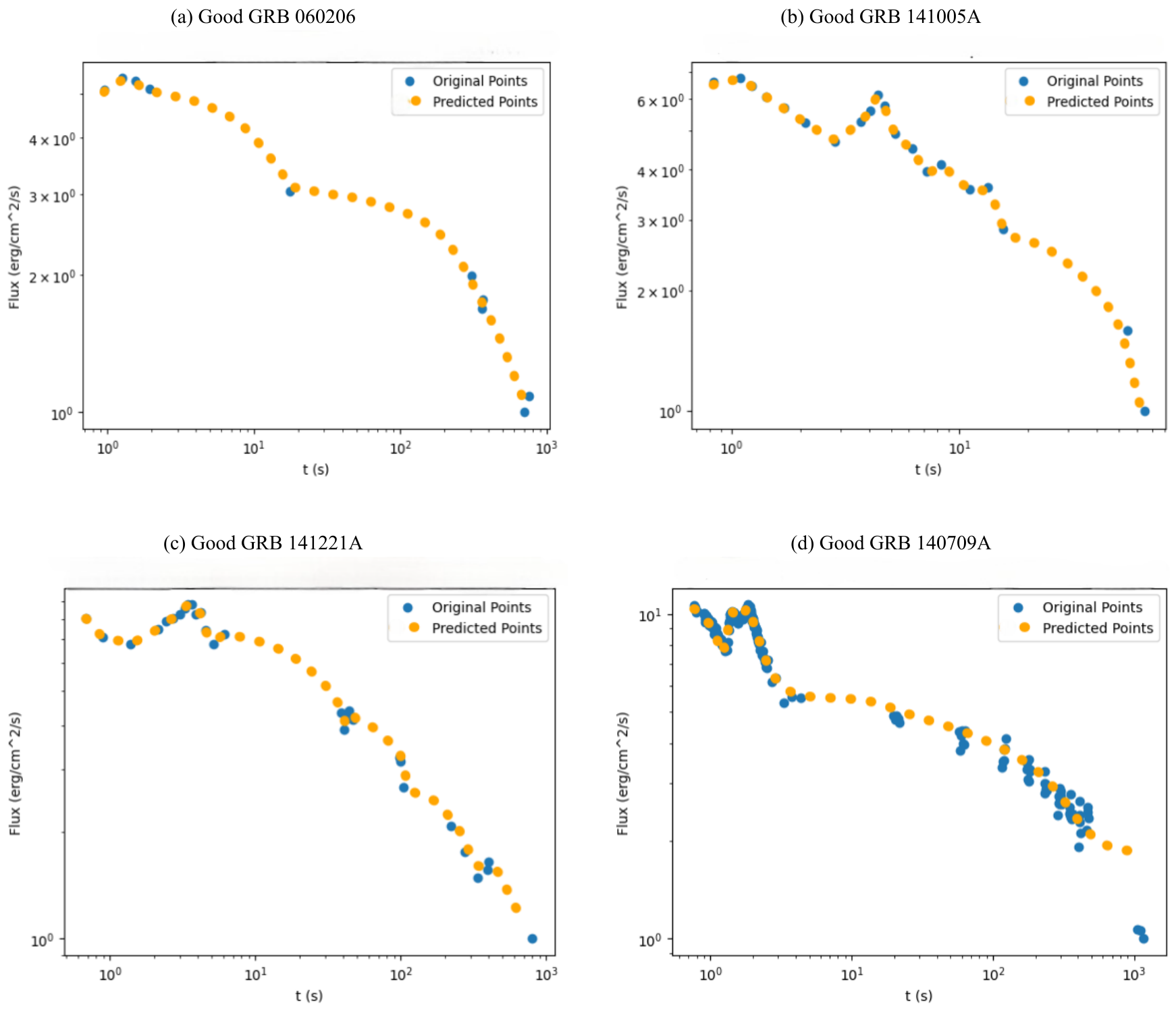}
    \caption{Bi-directional LSTM Light Curve Reconstruction for Bump Flare GRBs.}
    \label{fig: Break Bump GRBs3}
\end{figure*}

\begin{figure*}
    \centering
    \includegraphics[scale=0.6]{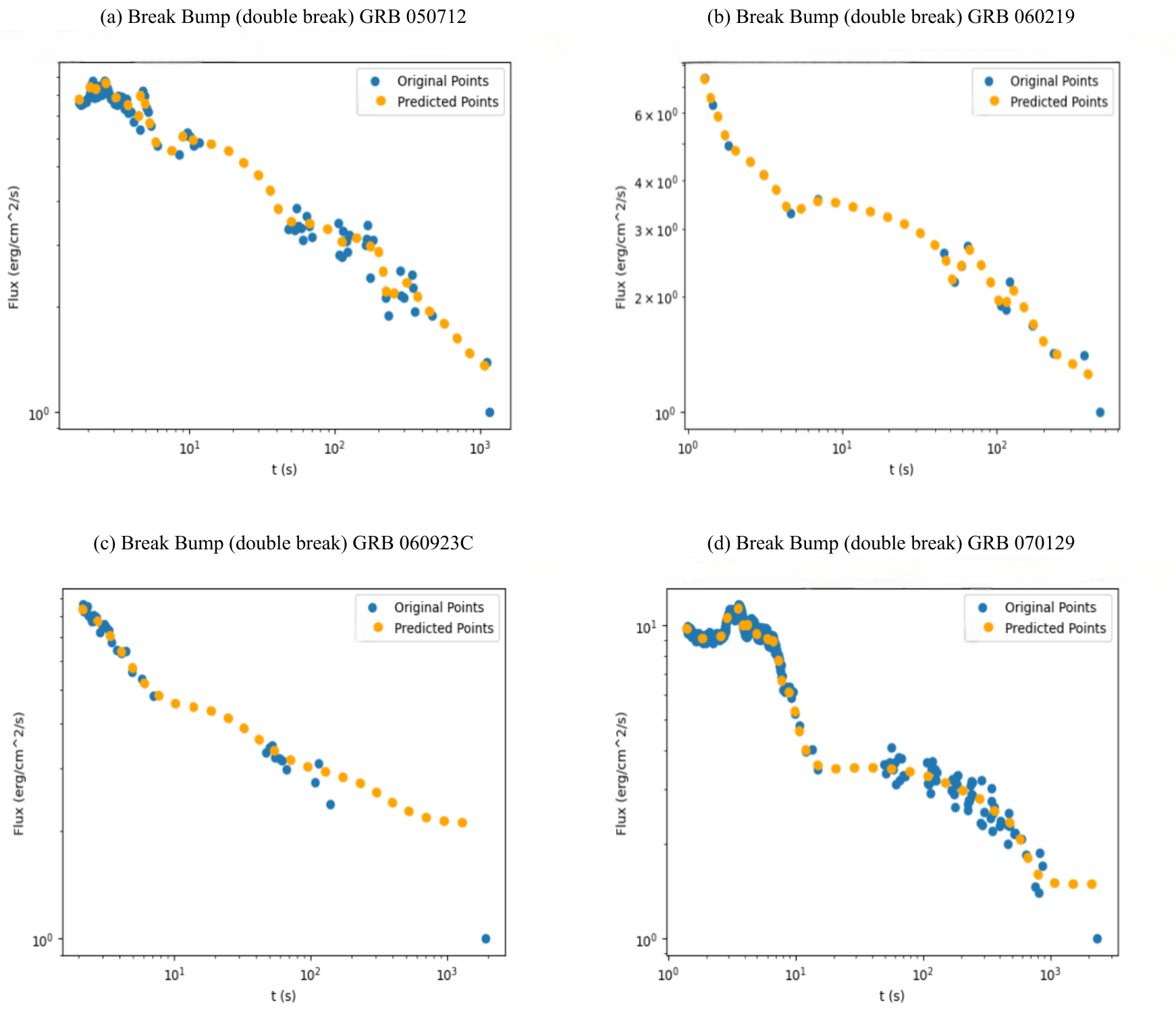}
    \caption{Bi-directional LSTM Light Curve Reconstruction for Break Bump (with double breaks) GRBs.}
    \label{fig: Break Bump GRBs4}
\end{figure*}

\begin{table*}\label{table 1}

	\centering
	\caption{Default size of Hyperparameters used while training our Bi-LSTM framework.}
	\label{table:1}
	\begin{tabular}{lc} 
		\hline
		Hyperparameter & Default Size\\
		\hline
            Batch Size & 900\\
            Maximum Number of Batches & 3 \\
		Minimum Number of Batches & 1 \\
	
		\hline
	\end{tabular}
\end{table*}

\begin{table*}
\centering
\caption{Comparison of $\%$ decrease in uncertainty over flux (after reconstruction) values in case of GRBs in good category for our BiLSTM model with W07, BPL, (W07+GP)  and (BPL+GP) models as done in \citep{2023}. Here GP denotes the Gaussian Process.}

\label{table:2}
\begin{tabular}{lccccr}
\hline
GRB ID & $\%EF_{\log_{10}F}$ (Bi-LSTM RC) & $\%EF_{\log_{10}F}$ (W07 RC)& $\%EF_{log_{10}F }$ (W07 + GP (RC)) & $\%EF_{\log_{10}F}$ (BPL RC) & $\%EF_{log_{10}F }$ (BPL + GP (RC)) \\ 
\hline
            050318  & $-14.19$ & $-24.05$ & $-27.68$ & $-23.11$ & $-24.84$\\
            050416A & $-23.07$ & $-33.54$ & $-34.04$ & $-22.5$ & $-20.55$\\
		050607  &  $-11.11$ & $-22.79$ & $-19.62$ & $-11.78$ & $-6.18$\\
            050712  & $-15.78$ & $-24.96$ & $-36.97$ & $-28.48$ & $-31.16$\\
            050713A & $-12.50$ & $-14.31$ & $-15.29$ & $-18.89$ & $-17.92$\\
            050822  & $-10.50$ & $-35.46$ & $-33.14$ & $-14.2$ & $-10.96$\\
            050824  & $-8.50$ & $-38.45$ & $-37.19$ & $-30.63$ & $-29.54$\\
             050826 & $-5.88$ & $-15.21$ & $-36.64$ & $-35.02$ & $-10$\\
            050915B & $-20.0$ & $-34.25$ & $-37.46$ & $-35.02$ & $-12.3$\\
            051016A & $-5.80$ & $-28.49$ & $-17.02$ & $-48.27$ & $-36.85$\\
            051109A & $-9.23$ & $-58.89$ & $-49.53$ & $-24.17$ & $-18.84$\\
            051221A & $-7.77$ & $-58.89$ & $-36.64$ & $-23.82$ & $-27.45$\\
            060105  & $-22.72$ & $-30.58$ & $-20.31$ & $-34.29$ & $-29.55$\\
             060108 & $-4.34$ & $-14.1$ & $-28.84$ & $-25.54$ & $-19.39$\\
             060109 & $-5.85$ & $-45.09$ & $-44.23$ & $-9.85$ & $-1$\\
            060124  & $-13.80$ & $-29.34$ & $-34.36$ & $-16.28$ & $-14.13$\\
             060218 & $-4.45$ & $-63$ & $-65.62$ & $-21.99$ & $-15.08$\\
             060306 & $-18.1$ & $-28.7$ & $-26.66$ & $-78.27$ & $-79.69$\\
            060418 & $-6.66$ & $-32.48$ & $-37.74$ & $-57.75$ & $-55.91$\\
            060421 & $-10.71$ & $-40.58$ & $-20.75$ & $-59.92$ & $-46.22$\\

\hline
\end{tabular}
\hfill

\end{table*}

\begin{table*}\label{tabel 3}
	\centering
	\caption{Table showing Error Fraction and $\%$ decrease in uncertainty over flux values after reconstruction of GRB LCs with our BiLSTM model for GRBs belonging to the Bump Flare, Break Bump (with single break), and Break Bump (with double break) category.}
	\label{table:5}
	\begin{tabular}{lcr} 
		\hline
		GRB ID &  $EF_{log_{10}F}$ (Bi-LSTM) & $\% EF_{log_{10}F}$ (Bi-LSTM RC) \\
		\hline
            050803  & $0.013$ & $-23.52$\\
            140304A  & $0.009$ & $-30.76$ \\
            140713A  & $0.007$ & $-30.0$\\
            190607B  & $0.0112$  & $-7.96$\\
            060206   & $0.013$ & $-23.52$\\
            141005A  & $0.019$ & $-19.83$\\
            141221A  & $0.0153$ & $-8.92$\\
            140709A  & $0.006$ & $-45.45$\\
            050712   & $0.005$ & $-11.83$ \\
            060219   & $0.015$   & $-22.68$\\
            060923C  & $0.008$   & $-38.67$ \\
            070129   & $0.006$   & $-25.68$\\
	
		\hline
	\end{tabular}
\end{table*}

\section{Introduction}
Gamma-ray bursts (GRBs) are astronomical events that can be detected at high redshifts \citep{2009Natur.461.1254T,2009arXiv0906.1578S,2011arXiv1105.4915c}. They emit radiation at various wavelengths, including gamma rays, X-rays, visible light, and radio waves. These bursts are exciting because of their intense brightness, allowing scientists to detect them from great distances. Since 2004, the Neil Gehrels Swift Observatory (Swift) has observed many gamma-ray bursts (GRB) at varying redshifts, luminosities, and durations \citep{2007RSPTA.365.1179O,2008A&A...484..293Z,2008AIPC.1065...67H,2020ApJ...893...77W}. The categorization scheme for short and long gamma-ray bursts (GRBs) was primarily developed using the $T_{90}$ parameter as a key factor, where $T_{90}$ refers to the period encompassing 5-95 $\%$ of the total burst fluence \citep{1982SvAL....8..354G, 1993ApJ...413L.101K}(\citep{1989Natur.340..126E,1993ApJ...413L.101K}). The prevalent perspective regarding the source of GRBs suggests that the extended Gamma-Ray Bursts (lGRBs) last for more than 2 seconds $(T_{90} > 2s)$ \citep{1993ApJ...405..273W,1998ApJ...494L..45P,1999Natur.401..453B,Wheeler:1999bu,2003ApJ...599..394M}. In contrast, short gamma-ray bursts (SGRBs) \cite{2015JHEAp...7...64B} have a $T_{90}$ of $\leq 2s$ \citep{1992ApJ...392L...9D,1992ApJ...395L..83N,Usov:1992zd,10.1093/mnras/270.3.480}.
Astrophysicists face several challenging obstacles in the study of Gamma-Ray Bursts (GRBs). These include the absence of a comprehensive classification system for GRBs \citep{10.1093/mnras/sts547,2016ApJ...832..136R,2017Ap&SS.362...70K}, incomplete data on redshift information for observed GRBs \citep{2007RSPTA.365.1179O,2012ApJ...745..168L,10.1093/mnras/sts020}, and gaps in the recorded light curves (LCs) \cite{2023arXiv230512126D}. This study addresses one of these challenges, specifically focusing on resolving the temporal gaps present in the LCs of GRBs.
The Neil Gehrels Swift Observatory (Swift, \cite{2004ApJ...611.1005G}) is crucial for observing GRB temporal properties. The Swift Burst Alert Telescope (BAT, $15-150$ keV, \cite{2005SSRv..120..143B}) detects prompt emission and the follow-up afterglow is detected using the X-ray (XRT, $0.3-10$ keV, \cite{2005SSRv..120..165B} and Ultra-Violet telescopes (UVOT $170 - 600$ nm), \cite{2005SSRv..120...95R}. In the first 15 years of its working \cite{2023Univ....9..113O}, Swift \cite{2007A&A...469..379E} has documented over $1000$ GRBs \cite{lien2016third} . Furthermore, due to rapid afterglow follow-up in several wavelengths, swift data has shown new features in the GRB LCs \citep{2005Natur.436..985T, 2006ApJ...642..389N, 2007ApJ...665..599T}.
Numerous X-Ray LCs frequently display a rapid decline in brightness after the initial emission, subsides at times, they are accompanied by a flare and/or plateau \citep{2006ApJ...642..354Z, 2006ApJ...642..389N, 2006ApJ...647.1213O, liang2007comprehensive, 2007ApJ...662.1093W,10.1111/j.1745-3933.2008.00560.x, 2010ApJ...722L.215D, 2016ApJ...825L..20D, 2017A&A...600A..98D}. XRT detections are available for $81\%$ of SWIFT GRBs, with $42\%$ consisting of X-Ray plateaus \citep{10.1111/j.1365-2966.2009.14913.x,2018ApJS..234...26L}. The X-ray plateau period generally spans from $10^2$ to $10^5$ seconds \citep{2019ApJS..245....1T,2020MNRAS.492.2847B,2021ApJ...922..102H} and is followed by a phase in which the decay follows a power-law pattern \cite{2022ApJS..261...25D}. Furthermore, approximately $30\%$ of optical LCs \cite{2023arXiv230512126D}, as observed by UVOT and ground-based facilities, show a gradual decline phase. \citep{2005Natur.435..178V, 2006ApJ...641..993K, 2006ApJ...637..889Z,2008MNRAS.387..497P, 2010ApJ...720.1513K, 2011MNRAS.414.3537P, 2011ApJ...734...96K, 2012ApJ...758...27L,2012MNRAS.426L..86O,2013A&A...557A..12Z,10.1093/mnras/sts066,2014A&A...565A..72M,2015ApJ...805...13L, 2018ApJ...863...50S,2018ApJS..234...26L, 2020ApJ...905L..26D}. This plateau can be fitted with a broken power-law (BPL) \citep{2006ApJ...642..354Z, 2009ApJ...698...43R}, or a smoothly broken power-law (SBPL) \cite{2007ApJ...662.1093W}. \\\\
The magnetar model often explains the plateau phenomenon, which is primarily based on dipole radiation from the energy transfer of a new neutron star (NS) \citep{2015JHEAp...7...64B,2014NuPhA.921...96M}. According to the model, the plateau ends when NS reaches the critical spin-down timescale. The uncertainty at time $T_{a}$ can be attributed to the uncertainty in the magnetar rotation period and magnetic field \citep{lyons2010can,corsi2009gamma}. However, the characterization of the plateau emission can be hindered by temporal gaps which can occur in the beginning, during, or at the end of the plateau. These may arise from the orbital period of satellites, lack of fast follow-up studies, atmospheric turbulence, and instrument failures \cite{2023arXiv230512126D}. Therefore, characterizing the light curve of gamma-ray bursts (GRBs) remains an important issue.
In terms of its physical structure, the plateau in diverse gamma-ray bursts (GRBs) exhibits more common characteristics, such as shallow decay phase \cite{2017ApJ...848...88D}, and prompt properties \citep{2019ApJS..245....1T,beniamini2020x,dereli2022wind}. These plateau features have garnered interest because they can be used to establish meaningful connections with plateau parameters and as tools for studying cosmology. \citep{10.1111/j.1745-3933.2008.00560.x,2010ApJ...722L.215D,2011ApJ...730..135D,dainotti2013determination, dainotti2015selection,dainotti2017study, li2018constraining} investigated the luminosity at the end of the plateau, ($L_{X,a}$) vs. its rest-frame time ($T^{*}_{X,a}$) (known as the Dainotti relation or 2D L-T relation) \cite{2010ApJ...722L.215D}. The optical plateau emissions have also revealed the presence of a two-dimensional relationship.
The plateau parameters are defined due to their numerous applications in cosmology \cite{2021ApJ...920..135X}. It is crucial to reconstruct light curves (LCs) that contain plateaus within designated plateau region. LCs that contain gaps are often unsuitable for cosmological purposes \cite{panaitescu2005models}. Furthermore, LCs with temporal gaps cannot be relied upon to accurately test theoretical models, such as the standard fireball model, which aims to explain gamma-ray burst (GRB) emissions \citep{1978MNRAS.183..359C,1993ApJ...405..278M,1999PhR...314..575P,Panaitescu:2000bk,10.1143/PTPS.143.33,2002ApJ...581.1236Z,2004RvMP...76.1143P, 2004IJMPA..19.2385Z,2006ApJ...642..354Z,2020ApJ...895...90S}. It was shown in \cite{2018ApJ...869..155S} that because there are gaps in the time data of certain light curves, it becomes difficult to determine the exact values of parameters such as the magnetic field, spin period, and electron energy fraction.\\\\
A lot of groups have previously evaluated LC morphology using interpolation \cite{lin2021bayesian}, deconvolution \cite{bright2023precise}, Simple Power Law (SPL) \cite{2010ApJ...711..641C}, Broken Power Law (BPL) \citep{2006ApJ...642..389N, 10.1111/j.1365-2966.2008.13990.x,10.1111/j.1365-2966.2009.14913.x, 2009ApJ...698...43R, 2009ApJ...707..328L}, smooth BPL \cite{1999ApL&C..39..281R}, W07 model \cite{2007ApJ...662.1093W} and Gaussian Processes \cite{2023arXiv230512126D}. The SPL model assumes a uniform power-law behavior for the entire duration of a Gamma-Ray Burst (GRB). Nonetheless, GRB light curves frequently display intricate and multi-component structures encompassing phases like prompt emission, plateau, and afterglow. Consequently, the SPL model cannot adequately capture these diverse features, resulting in a limited depiction of the underlying physical processes. On the other hand, the BPL determines $T_{a}, F_{a}$ and slope of the LC during the plateau ($\alpha_1$) and after the plateau ($\alpha_2$). This model allows for different slopes before and after the breakpoint, accommodating the characteristic steep rise and subsequent decay observed in GRB light curves.  While using BPL for fitting, there is a difference in the true value of the model parameters and the ones estimated using power law fitting. It has been observed that the slope of the electron energy distribution is overestimated based on the slope of the light curve before the break, but it is underestimated after the break. Hence, one must be careful when using Power-Laws for GRB LC fitting, whereas in smooth BPL, the size of the turnover spectral range $\Delta E$ \cite{1999ApL&C..39..281R} cannot be constrained, making it difficult to understand LCs. On the contrary, the W07 model determines the time at the end of the plateau ($T_{a}$), associated flux ($F_{a}$), and the temporal index after the plateau ($\alpha_{a}$). Most shallow decay segments for optical emission have a slope steeper than 0, while the Willingale function can only handle a true plateau (the slope before the break is 0) \cite{2007ApJ...662.1093W}. Hence, using the Willingale function to fit afterglow curves, especially optical ones, can sometimes be misleading. On the other hand, the method encompassing Gaussian Processes \cite{aigrain2022gaussian,liu2019gaussian,wang2022local} is limited to a particular set of GRBs. As most of the previous studies have some or the other limitations, in this manuscript, we propose a Bi-directional LSTM model that performs better than the existing models and can be generalized to any class of GRBs. \\\\
In recent years, Recurrent Neural Network (RNN) \cite{2020PhyD..40432306S} has emerged as a powerful tool for time series analysis. To count for drawbacks of traditional recurrent neural networks such as vanishing and exploding gradients that affects the long-term dependencies, algorithm such as LSTM \citep{10.1162/neco.1997.9.8.1735} were proposed. This vanilla version of the algorithm was further modified by adding an additional LSTM layer to understand the data flow from both the directions resulting into better understanding of the sequential patterns in the data and has been popularly called as bi-directional long short-term memory (BiLSTM) networks. A Bidirectional Long-Term Memory (BiLSTM) model \citep{schuster1997bidirectional,graves2005bidirectional,graves2005framewise,graves2013hybrid,huang2015bidirectional} involves dividing the state neurons into two parts: one responsible for the positive time direction (forward states) and another for the negative time direction (backward states). This bidirectional processing allows the model to capture dependencies from past and future contexts, enabling it to understand the input sequence more comprehensively. The BiLSTM model can take an input sequence, a sequence of words, time-series data, or any other sequential data and produce an output sequence that captures the learned representations and predictions based on the given input sequence and predicts future values based on historical sequences. Each element in the sequence corresponds to a specific time step or position. A BiLSTM consists of two LSTM layers: a forward LSTM and a backward LSTM. The forward LSTM processes the input sequence in a regular forward manner.
In contrast, the backward LSTM processes the sequence in reverse order, starting from the end and moving toward the beginning. Combining the outputs of both LSTMs, a BiLSTM can leverage information from past and future contexts, improving its prediction capabilities. Once the forward and backward LSTMs have processed the entire sequence, the hidden states from both LSTMs at each time step are concatenated. This creates a combined representation that captures the context from past and future positions relative to each time step. The concatenated hidden states are then passed through a fully connected layer or any suitable output layer for the specific task. Depending on the application, the output layer can perform classification \citep{liu2019bidirectional,li2020bidirectional,wang2021intelligent}, regression \citep{kaselimi2019bayesian,prabhudesai2019automatic,jiao2021non}, or sequence generation tasks \citep{nakamura2018outfit,mangal2019lstm,mootha2020stock}. During the inference or prediction phase, a BiLSTM model takes a new input sequence and performs forward and backward processing to capture the dependencies in both directions. This allows the model to make predictions by comprehensively understanding the entire sequence. \citep{9005997,kim2019bilstm,app10175841}. Our approach to reconstruction provides us with a realistic idea of the data that is likely to have existed in those missing sections, thus enhancing the overall density distribution of the light curves (LCs) across time. Consequently, this enables us to enhance the utility of LCs as standard candles and for theoretical modelling purposes.

\section{Methodology}\label{section : Methodology}
\subsection{Data Collection, Description, Pre-processing}
We use GRB light-curve data for all four categories of GRBs: Good GRBs, Break Bump GRBs (with single and double breaks), and Bump Flares GRBs \cite{2023arXiv230512126D}. The dataset also contains error values for these parameters. The light curve data of GRBs belonging to these classes generally have a small number of data points ($\approx 50 \text{ to } 100$). Such a small number of data points makes the problem of reconstructing the complete light curve much more challenging, as data turns out to be insufficient for training to produce any valid results. To overcome this challenge, we up-sample the light curve data in two steps or phases. The former includes the creation of new data points. At the same time, the latter duplicates data points, again and again, to correctly highlight the representation of each of the regions of the curve. 

\par In the first phase, we sort the dataset with respect to time to have a proper time sequence. We then use our BiLSTM model on the original dataset to understand the underlying pattern between the original data points. Then using this underlying pattern, we upsampled our dataset by creating ten new points between two neighboring data points in the original dataset. These data points were created in the original scale to reduce the chances of error and ambiguity. When dealing with transient events like GRBs, it's essential to observe the positive and negative fluctuations without losing detail. As a purely logarithmic scale only allows only positive values, hence to address this issue, the data was converted to a symmetric logarithmic scale using the mathematical function symlog expressed as:

\begin{align*} 
    \centering
    F(x) = ln\biggl(\frac{x}{a}\biggl) + 1
\end{align*} 

where $a$ denotes the minimum flux value, and $x$ denotes the flux values. This enabled our BiLSTM model to train better and give accurate predictions for the missing portions of the light curve. As the next step, we created multiple batches with batch size as an input parameter to analyze the effect of different batch sizes in the model's training. The optimal batch size based on experiments has been found to be 900 (table \ref{table:1}). Each of these batches is up-sampled to make several points sufficient enough for training. Since our method is based on localized point fitting, we first segregate the batch and then upsample them to avoid the same point being present in multiple batches.

\subsection{Model Architecture}
We propose a multi-BiLSTM framework with four BiLSTM layers and an additional dense layer (responsible for mapping the learned representation of the predictions). Instead of a single BiLSTM layer, the architecture uses a combination of multiple BiLSTM layers, each trained on data spanning distinct time stamp intervals of the GRB curve. Each model tries to focus and fit specific data regions instead of fitting on an entire dataset. The approach is based on localized fitting, which in our case, is better than any other technique so far in predicting the flux values of different classes of GRBs. There are 100 hidden units in each layer. 
Given that we work with upsampled data and engage in batch-wise training, as outlined in \cite{gasteiger2022influence}, each batch represents a randomized subset of the upsampled dataset. As a result, the gradients computed on different batches may vary, introducing noise in estimated gradients. To tackle this issue, we use Adam Optimizer \cite{bae2019does,mehta2019cnn}. The Adam optimizer adaptively adjusts the learning rate for each parameter based on the past gradients and squared gradients. We make use of Relu as an activation function in the dense layers so that our model can learn the non-linear dependency \cite{2020arXiv201007359K} between flux and time. 

\par The model's uniqueness is observed in the approach of its predictions. Previous models, like the Broken Power Law (BPL) model \cite{2023arXiv230512126D}, assume that a simple mathematical function can accurately describe the light curve with a few parameters. However, GRB light curves can exhibit complex and diverse behaviour, which a BPL may not be able to capture fully. Moreover, GRB light curves contain intricate features and variations at short timescales, which may be challenging to represent accurately with the Broken Power Law model. In contrast, the Bidirectional LSTM model captures complex temporal patterns in the light curve data more flexibly. Firstly, it can learn and model the dependencies between different time steps, enabling a more accurate representation of the underlying behaviour. Secondly, it can effectively model the missing GRB points in the light curve.

\subsection{Training and Validation}
The proposed method makes use of Keras and TensorFlow libraries \citep{Charles2013,2016arXiv160508695A} for its implementation. The training used an upsampled dataset of the GRBs. Every GRB goes through this training process individually, and the trained BiLSTM model is used to reconstruct the light curve for the same GRB. The entire dataset is split into $70\%$ training data and the rest $30\%$ for validation. The purpose of the validation dataset is to fine-tune the hyperparameters, ensuring an unbiased selection of optimal values.  
\par In general, the training procedure for BiLSTM can is summarised below: \vspace{0.2cm}

Bidirectional Long Short-Term Memory (BiLSTM) is an advanced version of the traditional Long Short-Term Memory (LSTM) neural network architecture, designed to capture information from both past and future sequences. There is dual information flow - one processing the input sequence in the original order (forward LSTM) and the other processing the sequence in reverse order (backward LSTM). The outputs of these two LSTMs are combined at each time step, resulting in a richer representation that includes information from both directions. 

\begin{itemize}
    \item Forward LSTM: The BiLSTM model processes all the upsampled input data for each time slice within the range $1 \leq t \leq T$ to make predictions. This involves performing a forward pass for both forward states (starting from $t = 1$ and progressing to $t = T$) and backward states (starting from $t = T$ and moving back to $t = 1$). After this, the model performs forward passes for the output neurons and starts computing input gate activation \(i_t\) at time step \(t\). The input gate is used to control how much information from the current input \(x_t\) and the previous hidden state \(h_{t-1}\) should be allowed to be used into the current cell state \(c_t\). In LSTMs- the basic building block of BiLSTM, the input gate activation is obtained by applying a sigmoid activation function \(\sigma\) to a linear combination of the input and previous hidden state followed by adding a bias term \(b_i\):
    \begin{align*}
        i_t = \sigma(W_{ix}x_t + W_{ih}h_{t-1} + b_i)  \\
    \end{align*}
    Next at time step \(t\), the forget gate \(f_t\) is activated. The forget gate controls how much of the previous cell state \(c_{t-1}\) should be forgotten or retained. The forget gate activation is obtained by using a sigmoid activation function \(\sigma\) on a linear combination of the input and previous hidden state with the inclusion of bias term \(b_f\): 

    \begin{align*}
         f_t = \sigma(W_{fx}x_t + W_{fh}h_{t-1}  + b_f) \\
    \end{align*}
    
    Subsequently, the output gate \(o_t\) is activated at time step \(t\) using a sigmoid activation function \(\sigma\) . This is then applied to a linear combination of the input, hidden state, and current cell state, followed by adding a bias term \(b_o\):
    \begin{align*}
        o_t = \sigma(W_{ox}x_t + W_{oh}h_{t-1} + W_{oc}c_t + b_o) \\ 
    \end{align*}
    
    The calculation of the candidate cell state \(\tilde{c}_t\) at time step \(t\) in a Bidirectional LSTM (BiLSTM) involves applying the hyperbolic tangent function \(\tanh\) to a linear combination of the input and previous hidden state, along with a bias term \(b_c\).

    \begin{align*}
       \tilde{c}_t = \tanh(W_{cx}x_t + W_{ch}h_{t-1} + b_c) \\ 
    \end{align*}
    
    The update of the cell state \(c_t\) at time step \(t\) in a BiLSTM is determined by adding the previous cell state \(c_{t-1}\) with the new candidate cell state \(\tilde{c}_t\). This is done by element-wise multiplication(Hadamard product) of the forget gate activation \(f_t\) and the previous cell state, and element-wise multiplication of the input gate activation \(i_t\) and the candidate cell state \(\tilde{c}_t\).

    \begin{align*}
        c_t = f_t \odot c_{t-1} + i_t \odot \tilde{c}_t
    \end{align*}

    Finally, the hidden state \(h_t\) at time step \(t\) is computed as the element-wise multiplication of the output gate activation \(o_t\) and the hyperbolic tangent of the updated cell state \(c_t\).
    
    \begin{align*}
       h_t = o_t \odot \tanh(c_t) \\ 
    \end{align*}

    In the given context, $x_t$ denotes the input at time step \(t\), \(h_t\) is the hidden state, \(c_t\) is the cell state, and $i_t$, $o_t$ and $f_t$ are the outcome of the input gate, output gate, and forget gate activations respectively.

\vspace{0.9em}

    \item Backward LSTM: In this step, the BiLSTM performs a backward pass over output neurons. The main difference here is the forward states are now from $t=T$ to $t=1$, and backward states start from $t=1$ to $t=T$. We then perform a backward pass from the forward to the backward states.
    \end{itemize}
   The backward pass involves three key steps: computing the gradient of the loss with respect to the output, propagating the gradient through the BiLSTM cell, and updating the parameters using the Adam optimizer. 

\par Continuing with this training process, we perform 100 iterations with our BiLSTM model with a batch size of 15. To enhance training efficiency, our approach adopts an early stopping criterion. This criterion will stop the training process if the model's validation loss doesn't decrease for a certain number of epochs. We also use persistence, which stops training if validation is still stable after a timeout. Early stopping prevents unnecessary computations when the actual accuracy is high, and there is not much change in it as the training process goes on.

 Given the regression nature of our study, the loss metric we consider is Mean Square Error (MSE). MSE \cite{allen1971mean} is a statistical metric that assesses a model's error level and is used as a loss function. It quantifies the mean of the squared deviations between the observed and expected values. A perfect model would exhibit an MSE of zero, indicating no error. As the model's error rate increases, the MSE value also increases. The MSE is also called the mean squared deviation (MSD). Mathematically :

 \begin{align*}
     \text{MSE} = \frac{1}{n} \sum_{i=1}^{n} (f_i - \hat{f}_i)^2 
 \end{align*}
 
 Where \textbf{$n$} is the total number of flux values in the data (after upsampling), $f_i$ denotes the actual (observed) flux value of the $i^{th}$ sample, and $\hat{f}_i$ denotes the predicted flux value for the $i^{th}$ sample. This method effectively reduces the risk of overfitting when the model is very specific to the training data and does not perform well on new data \citep{caruana2000overfitting,ying2019overview}. The training process was carried out several times to capture the hidden challenge of tracking the light curve direction; several of the interference-free equipment performances were studied to give the best results.

\section{Results}

In this study, we focus on proposing an algorithm for Light Curve reconstruction that focuses on small changes in the values of time stamps. This helps the model to achieve better results in addition to its applicability to a diverse set of GRBs

Applying our Bidirectional-LSTM procedure for LCR over GRBs, we see a reduction in the uncertainties on the flux values for the good GRBs as compared to the study made in \cite{2023arXiv230512126D}. As a first step towards computing the $\%$ decrease in the uncertainty associated with Flux, we use error fractions linked to the flux values for the original and reconstructed fit. The error fraction is computed using:
\begin{align*}
    \centering
    EF_{log_{10}F_{a}} = \bigg|\frac{\Delta log_{10}F_{a}}{log_{10}F_{a}}\bigg| 
\end{align*}
Here, $|\Delta log_{10}F_{a}|$ refers to the change in the value of flux just after the occurrence of afterglow with the initial fit and reconstructed fit.

For LCR, using the Gaussian Process, \cite{2023arXiv230512126D} chose the $95\%$ confidence interval. At any given time $t$, the expected flux values are computed by combining the adjusted flux value with noise derived from a random variable extracted from the Gaussian distribution of flux residuals. Further, they use the Gaussian Regressor function and perform MCMC simulations of the reconstructed LCs. After this, they pick the afterglow value to compute its uncertainty. For all the four models i.e W07, BPL, W07 with Gaussian Process (at $10\%$ and $20 \%$ noise level) and BPL with Gaussian Process (at $10\%$ and $20 \%$ noise level) the $\%$ decrease in flux uncertainty is calculated as follows:
\begin{equation}\label{equation 1}
    \% DEC = \frac{|EF_{X}^ {After}| - |EF_{X} ^{Before}|}{|EF_{X}^{Before}|} \times 100 
\end{equation}
To calculate the \% decrease in the uncertainty over the flux values, we make use of \ref{equation 1}. Our analysis shows that the $\%$ decrease in the uncertainty over flux values after reconstruction with our bidirectional LSTM model is lower than the W07, BPL,(W07 + GP) model (table \ref{table:2}). In the case of the (BPL + GP) model (table \ref{table:2}), our Bi-directional LSTM's decrease in uncertainty for flux is higher for 3 GRBs (GRB050607, GRB050915B, and 060109).

\section{Discussion and Conclusion}
 
Establishing connections between pertinent GRB variables is necessary to address these investigations while minimizing their associated uncertainties. Such missing points in the LC can significantly affect our understanding of collimating GRB jets, energy demands, and cosmological applications. Therefore, it is essential to ensure comprehensive monitoring of the light curves of GRBs. This work presents a BiLSTM model to reconstruct the light curves of GRBs (gamma-ray bursts). The Willingale function \cite{2007ApJ...662.1093W} is suitable for X-ray light curves but not optical ones, as it cannot handle steep slopes commonly observed in optical emissions. Another method introduced by \cite{2023arXiv230512126D} demonstrates LC reconstruction (LCR) solely for GRBs classified as "good." The aim of our work is to evaluate the reliability of the model's predictions and to reduce the uncertainty of flux results compared to these previous LCR studies.  Several tests were performed on the LC data, which led to the estimation of missing features in the original data. Compared to previous methods, our method provides higher accuracy in estimating results after flux results. Instead of generalizing some complex mathematical functions across all data, we approach analyzing some data and focusing on local areas to provide better results. Our model can predict the result of X-ray and optical light curves and can adjust to multi-wavelength GRB data, making it more convenient than Willingale \cite{2007ApJ...662.1093W} model, Broken Power Law Model, and the Gaussian processes \cite{ 2023arXiv230512126D} models.

Light curve reconstruction using BiLSTMs helps identify plateau characteristics in Gamma-Ray Burst (GRB) light curves that might go unnoticed. These objects can reveal progenitor processes and processes involved in  GRB emission. Gamma-ray bursts are often associated with non-thermal emission from relativistic jets. Understanding the structure and dynamics of these jets is vital to uncovering the bodies behind the GRBs. LCR reconstruction through this approach shows us how to utilize machine learning analysis incorporating improved and precise estimates of the plateau parameters; we can make predictions about the redshift information. This technique
allows us to determine the redshift of high-z GRBs (gamma-ray bursts) more effectively, thereby contributing to advancing studies related to Population III stars. The classification of GRBs based on their shape is also more precisely addressed by LCR. This new approach to the GRB LCR can help show the physical behavior of the explosion, describe processes in the light curve, and examine the evolution of the output at different times of the explosion. Such observations can reveal essential details of the underlying engine and mechanisms that control explosive energy release. GRBs are powerful cosmological tools because of their high luminosity and association with significant events. This method of generating light curves will help improve distance measurement and correlation between images of GRBs, thereby better limiting their cosmological parameters to understand the universe's expansion, the nature of dark energy, and the history of star formation.
 
With all these advantages, this bidirectional LSTM model has some limitations. Because these models require long-term inputs, the GRB light curve case's burst time and sampling rate may differ. Processing variable-length paths includes preprocessing steps such as padding or truncation, which can lead to artifacts or data loss. While BiLSTMs gather information from both preceding and subsequent steps, the scope of their window contents remains constrained. GRB light curves often show distinct and distinct patterns and prominent features may appear in the background. Long-term dependencies need to be captured correctly and result in data loss. In the future, we aim to reconstruct the GRB light curve using other deep-learning models. It was observed that models were learning best when the number of batches was set between 1 and 3, both inclusive. Our observation was consistent over all four categories of GRBs.

\par Subsequently, a sequence of BiLSTM models was trained, with each model consecutively targeting a distinct batch. The predictions made after the training of each model were collected similarly.
The error bars for the light curve were calculated using the standard deviation of the predictions for all data points over several iterations. Another important part of the data pre-processing was the time sequences were created for the model. This parameter was fixed to 1, i.e., each time sequence had a single point. This was done after trying out a lot of different values for this parameter. The predictions collected were then organized and converted back to the original scale from symlog (flux values) for plotting purposes.

\section{Data Availability}
All the data for the analysis has been collected from The Neil Gehrels SWIFT Observatory (SWIFT XRT) website $\href{https://www.swift.ac.uk/xrt_curves/00556753/}{https://www.swift.ac.uk/xrt_curves/00556753/}.$  

The initial values for calculating the error fraction have been adopted from
\cite{2020ApJ...903...18S}.

\bibliographystyle{mnras}
\bibliography{Bi-LSTM}

\bsp	
\label{lastpage}
\end{document}